%% file: ms.tex
\documentclass[letterpaper,twocolumn,10pt]{article}
\pdfoutput=1
\makeatletter
\newcommand\footnoteref[1]{\protected@xdef\@thefnmark{\ref{#1}}\@footnotemark}
\makeatother

\usepackage{array}
\newcolumntype{L}[1]{>{\raggedright\let\newline\\\arraybackslash\hspace{0pt}}m{#1}}
\newcolumntype{C}[1]{>{\centering\let\newline\\\arraybackslash\hspace{0pt}}m{#1}}
\newcolumntype{R}[1]{>{\raggedleft\let\newline\\\arraybackslash\hspace{0pt}}m{#1}}

\usepackage{usenix,epsfig,endnotes}

\usepackage[hyphens]{url}

\usepackage[OT1]{fontenc} 

\usepackage[table,usenames,dvipsnames]{xcolor}

\usepackage{algorithmic}
\usepackage{algorithm}

\usepackage{epsfig,endnotes}
\usepackage{epsfig}
\usepackage{amssymb}
\usepackage{amsmath}

\usepackage{booktabs}
\usepackage{colortbl}

\usepackage{caption}
\usepackage{subcaption}
\usepackage{multirow}

\usepackage{tabularx}
\usepackage{graphicx} 
\usepackage{color}
\usepackage{xspace}
\usepackage{thumbpdf}
\usepackage{hyperref}
\hypersetup{pdfstartview=FitH}
\usepackage{listings}
\usepackage{verbatim}
\usepackage{colortbl}

\usepackage{dirtree}
\usepackage{pbox}

\usepackage[shortcuts]{extdash}

\newcommand{\squeezeup}{\vspace{-2.5mm}}

\newenvironment{Itemize}%
{\begin{itemize}%
\setlength{\itemsep}{0pt}%
\setlength{\topsep}{0pt}%
\setlength{\partopsep}{0pt}%
\setlength{\parskip}{0pt}}%
{\end{itemize}}
{\begin{enumerate}%
\setlength{\itemsep}{0pt}%
\setlength{\topsep}{0pt}%
\setlength{\partopsep}{0pt}%
\setlength{\parskip}{0pt}}%
{\end{enumerate}}

\newcommand{\hide}[1]{}

\newcommand{\name}{Trevor\xspace}

\makeatletter
\newcommand\resetsubfigs{\setcounter{sub\@captype}{0}}
\makeatother

\begin{document}
\title{\name: Automatic configuration and scaling of stream processing pipelines}
\author{
Manu Bansal$^1$$^\dagger$\thanks{Work done while at Stanford} , Eyal Cidon$^2$$^\dagger$, Arjun Balasingam$^3$\footnotemark[1] , Aditya Gudipati$^1$, Christos Kozyrakis$^2$, Sachin Katti$^2$ \\
\vspace{5pt}
$^1$\normalsize{Uhana, Inc.} $^2$\normalsize{Stanford University} $^3$\normalsize{MIT}\\
$^{\dagger}$\normalsize{Co-primary authors} \\
}
\maketitle
\thispagestyle{plain}
\pagestyle{plain}
\input{abstract.tex} 
\input{intro.tex}

\input{motivation.tex}
\input{design.tex}
\input{implementation.tex}
\input{eval.tex}
\input{related.tex}

\input{conclusion.tex}

\bibliographystyle{abbrv} 
\bibliography{ms}
\end{document}

%% file: abstract.tex
\abstract{
Operating a distributed data stream processing workload efficiently at scale is
hard. The operator of the workload must parallelize and lay out tasks of the
workload with resources that match the requirement of target data rate. The
challenge is that neither the operator nor the programmer is typically aware of
the scaling behavior of the workload as a function of resources. An operator
manually searches for a safe operating point that can handle predicted peak
load and deploys with ample headroom for absorbing unpredictable spikes. Such
empirical, static over-provisioning is wasteful of both compute and human
resources.

We show that precise performance models can be automatically learned for
distributed stream processing systems that can predict the execution
performance of a job even before deployment. Further, those models can be used
to optimally schedule logically specified jobs onto available physical
hardware. Finally, those models and the derived execution schedules can be
refined online to dynamically adapt to unpredictable changes in the runtime
environment or auto-scale with variations in job load.\footnote{This publication is based on a previously published PhD thesis~\cite{manuthesis2018}}
}

%% file: intro.tex
\section{Introduction}
\label{intro}

\begin{figure}[t]
    \centering
    \begin{subfigure}[t]{0.15\textwidth}
    \centering
    \includegraphics[scale=0.50]{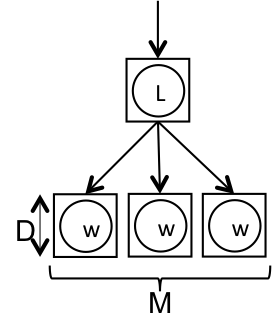}
    \caption{Configuring and scaling a webserver}
    \label{fig:webserver-scaling}
    \end{subfigure}
    \hfill 
    \begin{subfigure}[t]{0.30\textwidth}
    \centering
    \includegraphics[scale=0.50]{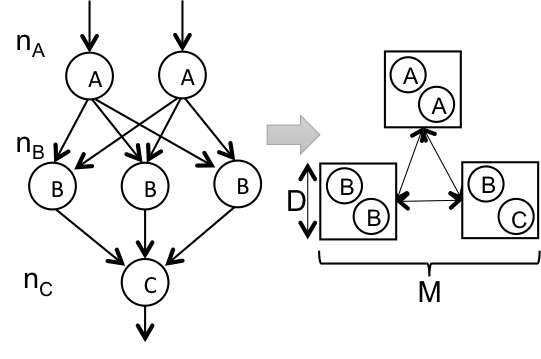}
    \caption{Configuring and scaling a stream processing pipeline}
    \label{fig:dag-scaling}
    \end{subfigure}
    \caption{Configuring and scaling webservers vs. stream
    processing pipelines} \label{fig:scaling-problem}
\end{figure}
\squeezeup

Real-time streaming workloads drive our favorite cloud services like Twitter
feed, LinkedIn profiles, Netflix personalized recommendations, Redfin
real-estate listings and mobile network control planes. However, those
workloads frequently run over-provisioned. They process millions of events per
minute with sub-second latency. In addition, they handle daily peak loads up to
3-5$\times$ the average load and up to 20$\times$ in case of special events
like a FIFA World Cup \cite{fifa, fifa-load}. Insufficient processing capacity
causes queue buildup and violation of the latency requirements. To guarantee
SLAs, operators generally deploy such workloads with fixed configurations for
peak load with headroom to absorb spikes \cite{twitter-fixed-provisioning} or
machine failures. An enterprise datacenter like that of Netflix can have 200
different streaming applications active over thousands of machines. At such
scale, fixed provisioning for peak-load can be extremely wasteful of resources.

\textbf{The configuration problem:} 
Real-time streaming workloads run over-provisioned due to the complexity of
\textit{configuration} - setting all the parallelism and physical resource
allocation parameters to guarantee processing of a certain target data rate
efficiently. For reference, we show a comparison with configuration parameters of 
cloud webservers in fig.
\ref{fig:scaling-problem}.  

A webserver can be configured for a certain query rate by setting virtual
machine dimensions D (CPU, memory and networking) and the number of virtual
machines M. A search over M is usually enough to find an efficient
configuration for a certain performance target. In contrast, configuring a
streaming workload DAG requires searching over a much larger space.  We must
set parallelism for each individual DAG node ($\lbrace n_X \rbrace$), machine
dimensions, and count. Then, we must pack DAG node instances onto the cluster
of machines.  Configuring the parallelism of a simple 3-stage DAG on a
100-machine cluster results in a search space of more than $7\times10^8$
combinations.\footnote{\label{note700m}$\sum_{1}^{100} (3M)^3$ where $M$ is the
number of machines used and we restrict the configuration space to allow only
up to 3 instances of a DAG node per machine}

In many cases, streaming applications are generated with higher-level DSLs like
Summingbird \cite{summingbird}. Lack of direct access to the runtime streaming
workload makes configuration even harder.  Moreover, the configuration must be
re-optimized as code changes or deployment datacenter changes because the
performance depends on both the application logic and the underlying
infrastructure.

\textbf{The scaling problem:} 
Another related but distinct problem in operating streaming workloads
efficiently is the complexity of {\em scaling} existing configurations to
higher loads. Unlike webservers, performance of a streaming workload does not
always grow with more parallelism, bigger containers, or more containers.
Replicating an existing configuration may even cause performance to drop due to
increase in communication overhead, as we show in sec.  \ref{sec:overview}.
Further, bottlenecks in the DAG shift as configurations are changed, requiring
relative parallelism of nodes to be adjusted for best performance. Those
dependencies require re-optimization of configuration every time target load
changes.

Reactive auto-scaling as in Dhalion \cite{dhalion} can be used to find
bottlenecks empirically and make point modifications to scale iteratively.
However, such a process can take hours to converge even for small DAGs, making
it impractical for responding to load spikes lasting few minutes
\cite{spikes-stats}. Moreover, the resulting configuration can be inefficient
due to limits on the extent of search space that can be explored empirically.
\\

\squeezeup
\begin{figure}[!h]
    \centering
    \begin{subfigure}[t]{0.22\textwidth}
    \centering
    \includegraphics[scale=0.45]{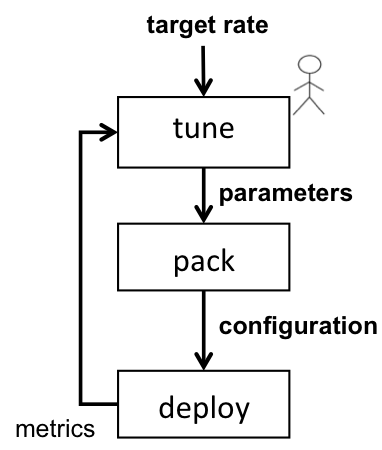}
    \caption{Current manual workflow}
    \label{fig:normal-workflow}
    \end{subfigure}
    \hfill 
    \begin{subfigure}[t]{0.24\textwidth}
    \centering
    \includegraphics[scale=0.45]{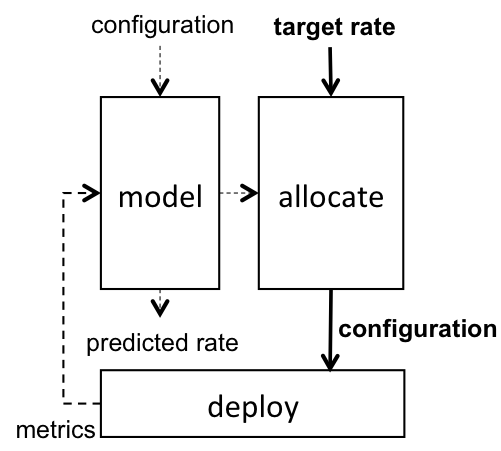}
    \caption{\name automated workflow}
    \label{fig:trevor-workflow}
    \end{subfigure}
    \caption{Manual vs. \name workflow for scaling stream processing pipelines}
    \label{fig:workflows}
\end{figure}
\squeezeup

In this paper, we present \name, a model-based solution to auto-configure and
auto-scale real-time streaming workloads. \name replaces the tedious manual
workflow of a hit-and-trial search process with an automated, single-shot
workflow as shown in fig. \ref{fig:workflows}. The core component of \name is
an {\em allocator} that takes in a target data processing rate and outputs a
configuration that will process the data rate efficiently.  The allocator
functions by training models that predict a workload's performance under any
given configuration.  We train those models as a one-time exercise for any
given workload. Models can be using runtime metrics collected from production
settings or test deployments. Once trained, the allocator can be used to
configure a workload for a fixed data rate or to repeatedly find new
configurations quickly to scale the workload as load changes.

We solve the challenge of modeling complex performance dependencies in the DAG
with the following insight: we can accurately predict the performance of a
configuration if we explicitly account for the cost of data communication
between DAG nodes in addition to the resource requirements (CPU, memory) for
each node.  We build on this insight to first develop prediction models and
then using them to design an allocator. 

In designing \name, we make the following concrete contributions:
\squeezeup
\begin{Itemize}
\item we show that a typical manually-tuned streaming workload configuration
can be 2-3$\times$ less resource-efficient than the optimal configuration for a
target data processing rate,
\item we design a performance model that can accurately predict the data rate that 
a parallelized stream processing DAG configuration can process, and
\item we present an allocation algorithm that quickly finds efficient physical
configurations for target data processing rates to auto-configure and auto-scale 
real-time streaming workloads.
\end{Itemize}

We implement \name as a stand-alone auto-configuration and auto-scaling agent
on top of the industry-standard Twitter Heron \cite{heron} stream processing
engine. We evaluate \name on a sample Word Count application, a Yahoo Ad
Analytics benchmark \cite{yahoo-streaming-benchmark} and a production mobile
network log processing pipeline. Our performance models are able to predict
workload data rates to within 10\% of measured rates for any configuration. Our
allocator is able to produce efficient configurations for given target rates
that operate within 10\% of the estimated optimal efficiency.

%% file: motivation.tex
\section{Background, Motivation and Insights}
\label{sec:overview}

Distributed stream processing systems like Twitter Heron \cite{heron}, Apache
Samza \cite{samza}, and Spark Streaming \cite{dstreams} provide map-reduce APIs
similar to batch processing systems like Hadoop and Spark. However, instead of
processing static batches of data, they operate continuously on unbounded
streams of data such as live Tweets, ad impressions and click events,
real-estate listings, or mobile user session events. The primary performance
metric of interest is the data rate (key-value tuples/sec or events/sec) that a
workload can process in real-time without building up queues or dropping data.
Typical applications expect sub-second processing delay per event. To
understand the need for and challenge in auto-configuration and auto-scaling,
we summarize the architecture of Twitter Heron as a reference stream processing
system and explore its performance sensitivity to varying configurations.

\subsection{Stream systems primer} 

\noindent \textbf{Programming model:} The basic structure of a Heron or Spark
Streaming distributed stream processing application is a directed acyclic graph
(DAG) of user-defined operations stitched together with map-reduce function
calls and connected with data grouping operators. DAG nodes run as parallel
instances over a set of physical machines, virtual machines, or containers with
fixed resources functioning as sandboxes in a framework like Mesos
\cite{mesos}.

\begin{figure}[!ht]
  \centering
  \begin{subfigure}[t]{0.18\textwidth}
    \centering
    \includegraphics[scale=0.45]{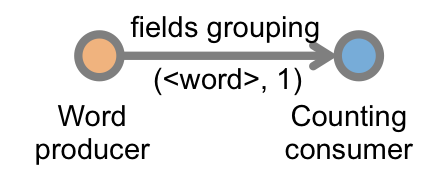}
    \caption{WordCount DAG} 
    \label{fig:wordcount}
  \end{subfigure}
  \hfill
  \begin{subfigure}[t]{0.29\textwidth}
    \centering
    \includegraphics[scale=0.5]{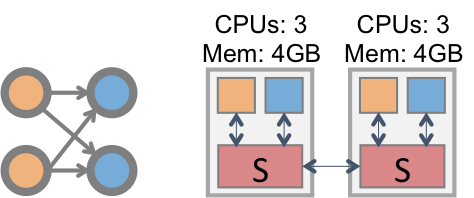}
    \caption{A specific configuration}
    \label{fig:wordcount-allocation}
  \end{subfigure}
  \caption{The WordCount sample application}
\end{figure}

A sample Heron WordCount application is shown in fig. \ref{fig:wordcount} with
two operations in the DAG. In this example, we are configuring the workload
with 2 parallel instances of each DAG node, as shown in fig.
\ref{fig:wordcount-allocation}.  The first node is a word-producer that emits
2-tuples with key being random words drawn from a finite vocabulary and value
being 1. The second node is a counting-consumer that maintains running counts of
word occurrences as a key-value store by adding up values of incoming tuples
grouped by key. 

Heron provides default implementations of 3 data grouping operators, which
use different mapping functions to decide the downstream instance(s) for each tuple:
fields-grouping(hash function), shuffle-grouping(random instance) and all-grouping(all instances).
In this scenario, the fields-grouping operator is used to ensures that all occurrences 
of a specific word are routed to the same instance of the counting-consumer.
\\

\noindent \textbf{Physical deployment:} A parallelized Heron DAG is deployed on
physical infrastructure using containers in a framework like Mesos. Containers
are sandboxed runtime environments that act as lightweight virtual machines.
Each container has a specific set of CPU and memory resources dedicated to it.
Containers are finally deployed onto physical or virtual machines in a
datacenter or cloud by a cluster scheduling framework such as Aurora or
Marathon. In the example shown, we are deploying the WordCount workload on 2
containers each with 3 CPUs and 4GB of memory packing DAG node instances into
the containers in a round-robin manner.

The Heron runtime system sets up each container with a process called the {\em
stream manager}, shown as S in fig. \ref{fig:wordcount-allocation}. Instances
packed into a container communicate with each other and remote instances
through the stream manager. The stream manager is the single point of data
entry and exit for any container. In general, a cluster of containers is a
fully-connected graph with edges between stream managers. Stream managers also
consume compute resources on containers.
\\

\noindent \textbf{Configuration:} Operators of a Heron or Spark streaming
workload must configure a set of runtime parameters before deploying it: 1)
parallelism of each DAG node (P), 2) dimensions (CPU and memory) (D) of
containers, 3) count of containers (M), and 4) packing of instances onto
containers (tab. \ref{table:control-parameters}). For best performance,
parallelism is configured for each DAG node separately. Similar to virtual
machines, containers are units of physical resources, however, they can be
sized on continuous axes. For example, it is possible to deploy containers with
2.5 CPUs and 3.1GB of memory. Packing of node instances onto containers affects
data locality and the overhead of data shuffling.

\begin{table}[!h]
\centering
\begin{tabular}{|p{2cm}|p{1.4cm}|p{3.0cm}|}
\hline
\footnotesize{\textbf{Heron}} & \footnotesize{\textbf{Spark}} & \footnotesize{\textbf{Role}}\\
\hline
\footnotesize{parallelism (P)} & \footnotesize{partitions} & \footnotesize{degree of DAG node parallelism} \\
\hline
\footnotesize{container dimension (D)} & \footnotesize{executor dimension} & \footnotesize{unit of resource allocation}\\
\hline
\footnotesize{container count (M)} & \footnotesize{executor count} & \footnotesize{scaling of resource allocation}\\
\hline
\footnotesize{packing algorithm} & \footnotesize{locality wait} & \footnotesize{exploiting data locality}\\
\hline
\end{tabular}
\caption{Key configuration parameters affecting
distributed streaming workload performance in Heron and Spark}
\label{table:control-parameters}
\end{table}
\squeezeup

As we show next, the performance of streaming workloads is very sensitive to
configuration. However, finding the right configuration is challenging.  A
simple 3-node application can be parallelized and packed onto a cluster of 100
virtual machines in more than 700 million combinations\footnoteref{note700m}.
The space becomes even bigger if we permit container dimensions to vary from
full machine size to a third-of-a-machine. In typical enterprise data-centers,
smaller containers get scheduled faster due to more available slots, making
container dimensions an important parameter to search over.  While both Heron
and Spark provide default configurations, they can be off by upto $3\times$
from ideal configurations.

\begin{figure*}[!ht]
    \begin{subfigure}[t]{0.32\textwidth}
        \includegraphics[scale=0.3]{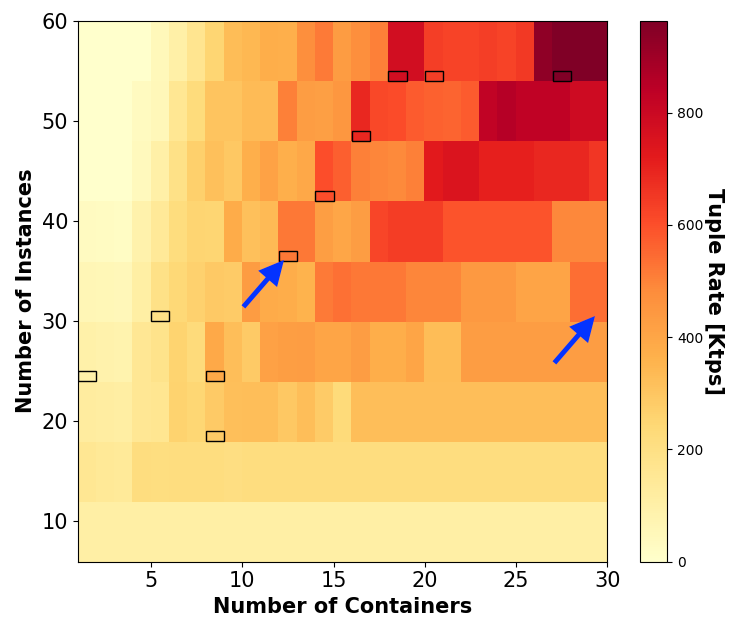}
        \caption{Heatmap of tuple rate for different tuning mixes}
        \label{fig:heatmap}
    \end{subfigure}
    \hfill
    \begin{subfigure}[t]{0.32\textwidth}
        \includegraphics[scale=0.29]{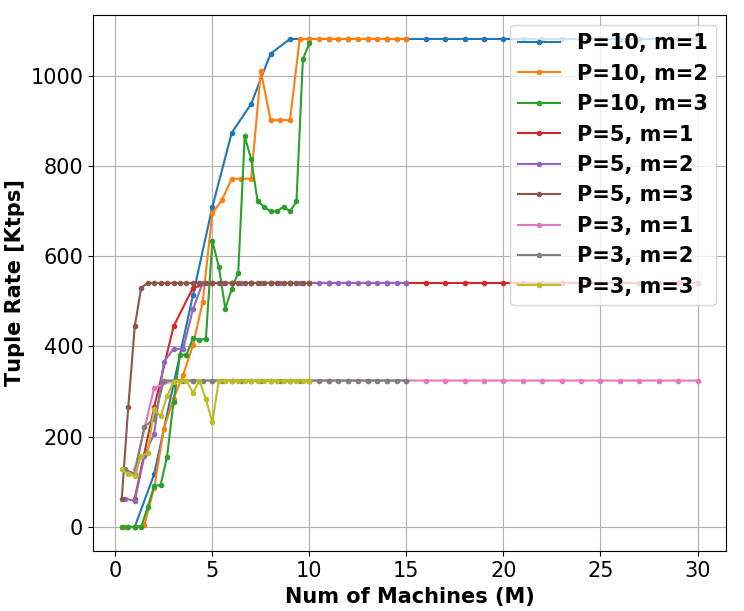}
        \caption{Effects of increasing number of worker nodes}
        \label{fig:machine_scale}
    \end{subfigure}
    \hfill
    \begin{subfigure}[t]{0.32\textwidth}
        \includegraphics[scale=0.3]{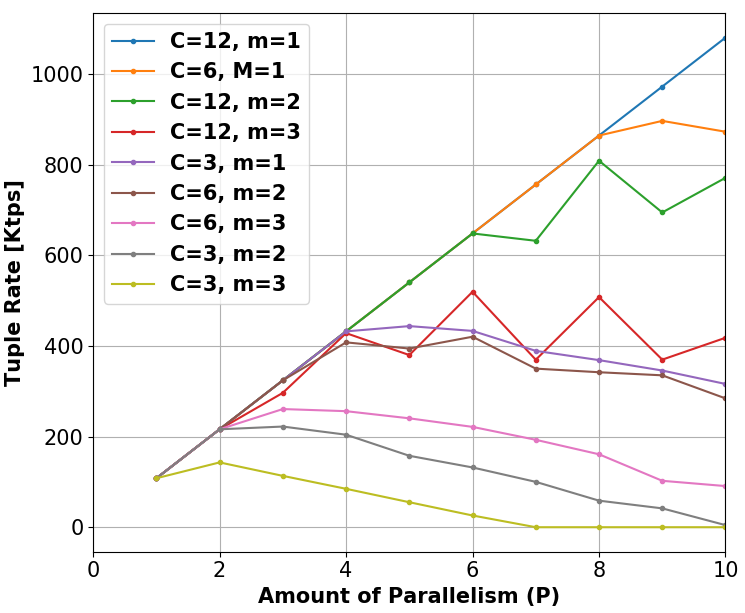}
        \caption{Effects of increasing number of node parallelism}
        \label{fig:par_scale}
    \end{subfigure}
    \caption{Configuration metrics and performance sensitivity: m - Multiplicity, P - Parallelism, C - \#Containers}
    \label{fig:yahooapp-sensitivity}
\end{figure*}
\squeezeup

\subsection{Performance sensitivity \& need for auto-configuring} 


\noindent {\bf WordCount workload:} We studied the performance sensitivity of
Heron workloads to configuration using the WordCount application in toy setups.
The most insightful results are listed in tab.
\ref{table:wordcount-performance}. A configuration is specified by parallelism
of the word-producer (\#W), parallelism of the counting-consumer (\#C), number
of containers (M), and the packing plan. We fixed container dimensions to 3
CPUs and 4GB memory. For each configuration, we measured the achieved
tuple-rate expressed in kilo-tuples-per-second (ktps). A packing (w,c)
$\leftrightarrow$ (w,c) represents two containers, each with an instance of
each DAG node.

\begin{table}[ht]
\centering
\footnotesize
\begin{tabular}{|p{0.15cm}||p{0.3cm}|p{0.3cm}|p{0.2cm}|p{1.85cm}|R{1.25cm}|p{0.85cm}|}
\hline
ID & \textbf{\#W} & \textbf{\#C} & \textbf{M} & \textbf{packing} & \textbf{rate(ktps)} & \textbf{bound}\\
\hline
1 & 1 & 1 & 2 & (w) $\rightarrow$ (c) & 658 & $\approx R_c$\\
\hline
2 & 2 & 2 & 2 & (w,c) $\leftrightarrow$ (w,c) & 965 & comm \\
\hline
3 & 2 & 2 & 2 & (w,w) $\rightarrow$ (c,c) & 648 & comm \\
\hline
4 & 1 & -1 & 2 & (w) $\rightarrow$ (0) & 839 & $\approx R_w$\\
\hline
5 & 1 & 2 & 3 & (w) $\rightarrow$ (c)$\times 2$ & 899 & $\approx R_w$\\
\hline
\hline
6 & 2 & 2 & 4 & (w)$\times 2$ $\rightarrow$ (c)$\times 2$ & 1319 & $2 \times R_c$\\
\hline
7 & 2 & 3 & 5 & (w)$\times 2$ $\rightarrow$ (c)$\times 3$ & 1779 & $2 \times R_w$\\
\hline
8 & 2 & 4 & 6 & (w)$\times 2$ $\rightarrow$ (c)$\times 4$ & 1847 & $2 \times R_w$\\
\hline
9 & 2 & 5 & 7 & (w)$\times 2$ $\rightarrow$ (c)$\times 5$ & 1582 & drop\\
\hline
\end{tabular}
\caption{WordCount performance under different configurations}
\label{table:wordcount-performance}
\end{table}
\squeezeup

We first tried to double the performance from a base configuration of one
instance each (ID=1) to two instances of each node (ID=2,3), also doubling
containers.  Any DAG instance could use only 1 CPU at the most due to
single-threaded implementation. With at most three instances per container
(count stream manager) in any configuration, we expected our configuration
switch to double performance. However, we only saw a $1.5\times$ increase from
658ktps to 965ktps. 

To isolate the performance bottleneck, we replaced the counting-consumer node C
with a null operation that would simply drop all incoming tuples. This resulted
in 839ktps (ID=4), higher than the base configuration. This indicated that C
was the slower node with peak processing rate $R_c$ of 658ktps and W had a peak
rate $R_w$ of 839ktps. With those numbers, we expected the ID=2 configuration
to process at $2 * R_c$, or twice the rate of ID=1; however, we realized that
the ID=2 configuration was bottlenecked on data shuffling. The permuted
configuration of ID=3 was even more shuffling limited due to all data crossing
containers. Inspecting resource utilization revealed that shuffling performance
was CPU-limited at stream managers in those configurations.


The next set of experiments were focused on effects of shuffling overhead.  We
successively increased the parallelism of the slower node C while adding
containers in proportion. Packing only one instance per container gave us the
most number of stream managers per instance and thus most shuffling capacity in
the configuration. As expected, this shifted the bottleneck from shuffling to
computation within a DAG node (ID=6,7,8).  This exploration revealed another
effect -- over-parallelization (ID=9) -- where excess instances of a
non-bottleneck node caused performance to drop.
\\

\begin{figure}[ht]
\squeezeup
    \centering
    \includegraphics[scale=0.25]{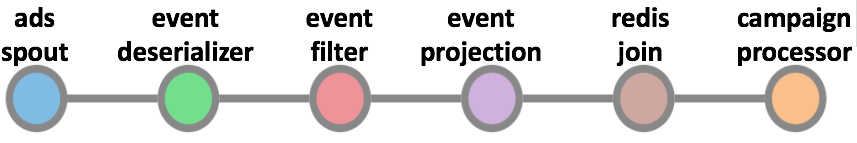}
    \caption{Yahoo ad-analytics processing
    benchmark\cite{yahoo-streaming-benchmark}}
    \label{fig:yahooapp}
\squeezeup
\end{figure}

\noindent {\bf AdAnalytics workload:} We further analyzed performance
sensitivity on a 6-node Yahoo ad-analytics pipeline
\cite{yahoo-streaming-benchmark} that serves as a real-world stream application
benchmark (fig. \ref{fig:yahooapp}).  Similar to WordCount, it showed
counter-intuitive performance sensitivity to configuration. 

In fig.
\ref{fig:heatmap} \footnote{ The results in fig. \ref{fig:yahooapp-sensitivity}
were generated by analyzing data flow through candidate configurations. Actual
measurements for a subset of those configurations showed even steeper
performance drop as parallelism was increased. We suspect that to be due to
instance execution interference caused by effects like excess context
switching.  }, we show a heatmap of the tuple rate under different mixes of
parallelism and number of containers. Parallel instances were packed into
containers in a round robin fashion.  For a few tuple-rates, we highlight the
optimal configurations out of the explored space with rectangles.  

We can see
that performance varies non-monotonically in the configuration space. There is
no clear path through the space we can traverse greedily to reach the optimal
configuration for a given target tuple rate. In fact, for the specific target
rate of 500ktps, even in this small subspace of configurations that an operator
might manually explore, the poorest configuration uses 30 containers as opposed
to 12 in the best configuration (blue arrows).  This amounts to a 2.5$\times$
efficiency cost of incomplete exploration. 

In fig. \ref{fig:machine_scale}, we show that simply adding machines while
fixing the parallelism will have limited benefit since we might become
bottlenecked on one of the replications of the nodes. P denotes the per-node
parallelism while M denotes {\em container multiplicity} which is the number of
containers per virtual machine. Higher multiplicity implies smaller containers.
In fig.  \ref{fig:par_scale}, we show that simply increasing the number of
instances also can lead to performance degradation, this is because adding
replications also increases the shuffling of tuples in the topology which adds
communication overhead. C denotes the number of total containers in the
configuration.
\\


\noindent {\bf Insights:} Based on the preceding sensitivity exploration, we
draw following insights that are key to the design of \name:
\squeezeup
\begin{Itemize}
\item Workloads can become shuffling-limited in certain configurations.
Shuffling limits could be hit due to CPU overhead, not just network resources.
\item Under-parallelization of DAG nodes can cause computation bottlenecks
while over-parallelization can cause excess data shuffling overhead.
\item Using few big containers can cause communication bottlenecks while using
many small containers can force excess parallelism and data shuffling overhead.
\end{Itemize}
In summary, it is essential to analyze the data shuffling overhead in addition
to DAG node computation cost to accurately predict the performance of a given
configuration.

\subsection{Load variation \& need for auto-scaling}

Streaming applications processing social networks or mobile networks need to
deal with significant load variations. LinkedIn processes 12.7 mil-events/s on
average over 24 hours with 18 mil-events/s at peak load; Netflix processes 4.6
mil-events/s daily average with peak of 8 mil-events/s \cite{load-numbers}. A
mobile network with 3000 cells processes 1.6k events/s to 83k events/s over a
week. In addition to diurnal and weekly variations, those services must handle
transient load peaks sometimes lasting only a few minutes with up to 25$\times$
higher load than average \cite{castle}. 

Static configurations to accommodate such variations are extremely wasteful of
resources. Dynamic auto-scaling using a system like Dhalion \cite{dhalion} can
potentially improve efficiency. However, using such a reactive system for
scaling complex DAGs can be too slow. Even for a simple 3-node WordCount DAG,
scaling the configuration from 1 mil-tuples/min to 4 mil-tuples/min took the
system more than 30 minutes. This convergence time becomes higher with more DAG
nodes.

To deal with transient load peaks over a high dynamic range, our goal in
designing \name is to enable computing efficient configurations for target
performance without requiring any manual intervention or hit-and-trial feedback
loops.

%% file: design.tex
\section{Design}
\label{sec:modeling}

\begin{figure}[!h]
\squeezeup
    \includegraphics[scale=0.50]{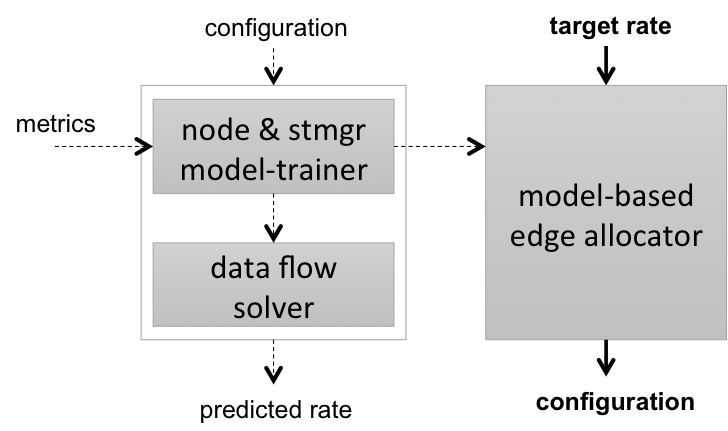}
    \caption{\name architecture}
    \label{fig:trevor-architecture}
\squeezeup
\end{figure}

\name meets the objective of auto-configuring and auto-scaling streaming
workloads by providing a {\em declarative allocator}. With \name, an operator
can simply declare a target tuple-rate of processing. In return, the allocator
will output an efficient configuration capable of sustaining the target
performance. This eliminates the need for a reactive, tedious and slow process
for finding efficient configurations empirically. While the reactive process
can take multiple iterations of tuning and deployment to hit a target
performance goal, the \name allocator finds the right configuration instantly.
This enables \name to auto-configure complex workloads over large clusters and
auto-scale them in response to load changes every minute. \name workflow and
system architecture are shown in fig. \ref{fig:trevor-architecture}.

In designing \name, we solve two key challenges. First, each workload has
different performance sensitivities to configuration parameters, specific to
the workload.  The ideal configuration for a target rate depends on the
computational load of user-defined DAG nodes. To find efficient configurations
for arbitrary workloads, we need to model workload-specific performance
behavior. Second, the configuration search space grows exponentially with DAG
size and deployment scale. We need a computationally efficient way to translate
workload performance models into configurations that can meet the declared
performance target while using resources efficiently.

We solve the design challenges with two key components: a modeling component
and a workload allocator. The modeling component uses runtime metrics data for
a workload to learn its application-specific performance behavior. Once
trained, the workload model can accurately predict the tuple-rate that the
workload will sustain under a given configuration. The models are used as input
to a workload allocation algorithm that produces an efficient configuration for
a declared performance target. The allocator sets DAG node parallelism,
container dimensions, container count and packing of instances to containers.
It functions by analyzing the DAG structure together with workload models to
compute efficient configurations without needing to search over the space of
configurations. The main insight that enables \name's design is that
computation and communication cost of data flow through a configuration must be
precisely accounted for. Both the modeling and the allocator components are
designed based on this insight.

\subsection{Workload models}

\subsubsection{DAG node models}

\begin{figure}[!h]
    \centering
    \includegraphics[scale=0.55]{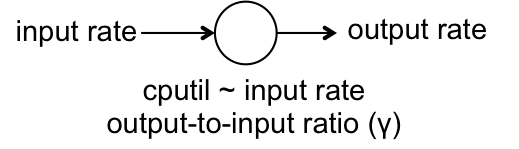}
    \caption{Basic DAG node model}
    \label{fig:basic-node-model}
\squeezeup
\end{figure}

The basic unit for application-specific resource modeling in \name is a DAG
node. For each DAG node, we learn 1) a relation M between CPU utilization
(cputil) and input tuple rate, and 2) the output-to-input ratio ($\gamma$) of
data rate for each DAG node, as shown in fig.  \ref{fig:basic-node-model}.
Per-node models ${M}$ completely characterize the computational footprint of
user-defined operations in the workload. In addition, output-to-input ratios 
specify the change to flow rate as data passes through user-defined operations 
like aggregation, filter and map functions that can decrease or increase the 
rate for downstream nodes. 

\begin{figure*}[!h]
    \centering
    \begin{subfigure}[t]{0.40\textwidth}
      \includegraphics[scale=0.28]{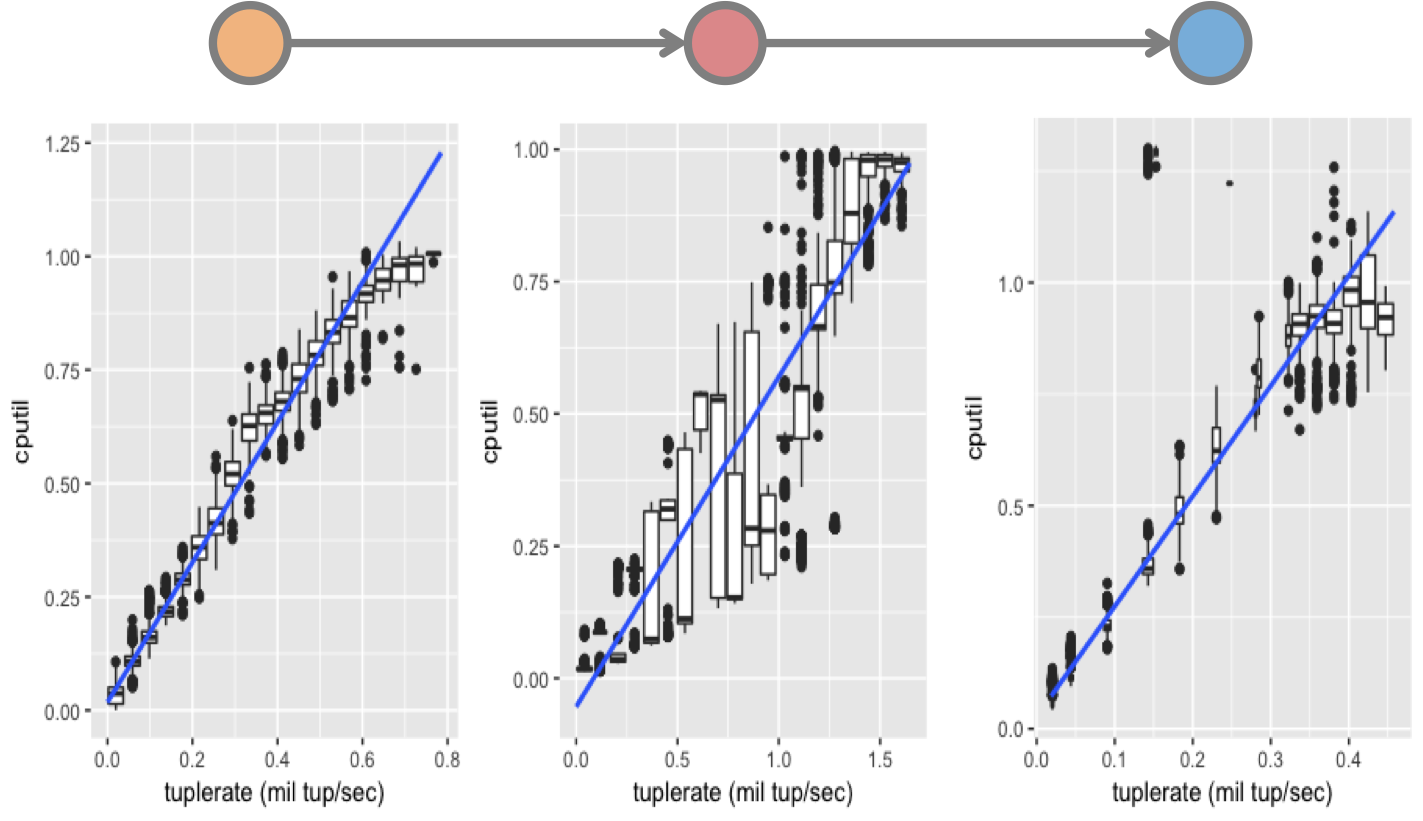}
      \caption{WordCount node CPU models} 
      \label{fig:wordcount_cpu}
    \end{subfigure}
    \begin{subfigure}[t]{0.28\textwidth}
        \centering
        \includegraphics[scale=0.23]{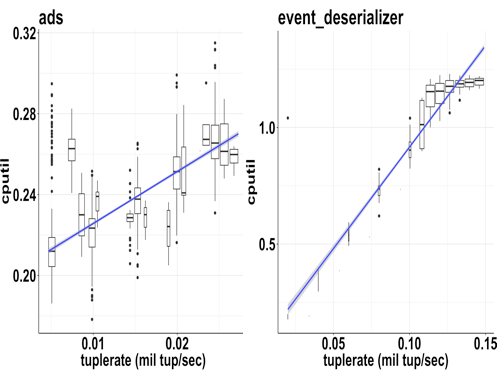}
        \caption{Two of the component cost models for AdAnalytics}
        \label{fig:models_yahoo}
    \end{subfigure}
    \begin{subfigure}[t]{0.28\textwidth}
      \centering
      \includegraphics[scale=0.2]{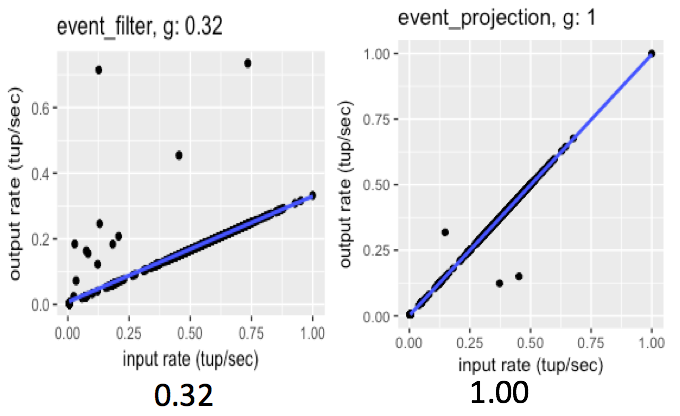}
      \caption{AdAnalytics node output-to-input ratios}
      \label{fig:inout}
    \end{subfigure}
    \caption{DAG node models}
    \label{fig:dag-node-models}
\end{figure*}

Models for DAG nodes capture the user-defined compute operations, however, we
we also need to model the data shuffling cost. Our performance sensitivity
exploration in sec. \ref{sec:overview} showed that workloads can become
performance limited due to the computation burden of data shuffling. Recall
that all data communication in Heron passes through the stream manager that is
inserted by the Heron runtime system on each container. We exploit this runtime
structure to model the compute cost of communication by treating the stream
manager as another node in the DAG that splits each edge in the
DAG.  With this transformation, the original WordCount DAG $W \rightarrow C$
becomes $W \rightarrow S \rightarrow C$, where S is a stream manager node. Then,
its compute cost can be modeled in the same way as user-defined nodes.

We train for the models ${M}$ using runtime CPU utilization and data flow rate
metrics. As the load varies naturally in the workload due to diurnal patterns
or spikes, metric timeseries span a range of operational loads that expose the
performance sensitivity. With this data, we fit linear models to node CPU
utilization. Examples of datasets and linear models for the WordCount
application are shown in fig. \ref{fig:wordcount_cpu} for emulated load
variation. We obtained a fit of 
$R^2$ = 0.7036, 0.7434, 0.7712\footnote{$R^2$ describes the portion of variance of the dependent variable (CPU) that is explained by the dependent variable (tuplerate), high $R^2$ means better model fit} for the three components W, S and C respectively. 
(Both W and C are single-threaded implementations but CPU utilizations exceed 1.0. 
This can happen due to additional supporting threads from the Heron runtime system.)

Node CPU utilization is not guaranteed to be a linear function of the input
data rate in general. However, we expect most design patterns to follow
close-to-linear relations. For example, a random number or string generator is
doing constant amount of work per input (or output) data tuple. Similarly, a
filter operation is usually evaluating a fixed-cost comparator per data tuple.
The stream manager as a node generally showed higher variance in CPU
utilization than user-defined operations, likely due to network interaction and
buffering effects. However, the predominant relation in the data showed a good
linear fit.  Our modeling of nodes does not require linear kernels but we have
found this to be a good approximation in practice. We evaluate the effects of
this choice in more detail later. 

We train for $\gamma$ also using a linear model. Assuming $\gamma$ is a
constant factor, we can recover it as the slope of a linear model between the
output data rate and the input data rate of the node under different loads. An
example of learning $\gamma$'s is shown in fig. \ref{fig:inout}. It shows two
nodes taken from the Yahoo AdAnalytics DAG. The {\tt event\_projection} node
takes in a stream of ad events and emits each event after modifying the data
representation. As expected, we recover $\gamma=1.0$ for this node. The {\tt
event\_filter} node passes about a third of the ad events coming in. Again, we
correctly recover $\gamma=0.32$. For the stream manager, $\gamma = 1$ by
definition as it is simply a router of data.

Our node models by themselves, are not sufficient to predict the
performance of a workload configuration.  As a case study, we tried to predict
the tuple rate of a base configuration of WordCount. The configuration used 2
instances each of W and C deployed on 2 containers with 3 CPUs each. The
packing was $(w,c) \leftrightarrow (w,c)$.  We first predicted the tuple rate
to be 1320ktps by saturating CPU utilization of W and C instances, however, the
measured performance was 965ktps. Next, we factored in the CPU utilization of S
too, however, we still predicted the same value which was 37\% higher than the
true value. This gap was due to inaccurate predictions of data rate being
processed by each node, including the stream manager. Next, we present the
design of a data flow solver that accurately predicts workload performance
using node models, thus also proving that our node models accurately capture
application specifics.


\subsubsection{Data flow solver}

The steady state data flow in a workload configuration depends on DAG node CPU
models, DAG edge structure, data shuffling operators on the edges and the
packing plan of node instances onto containers.  Completely modeling a
streaming workload requires the ability to analyze and predict the data flow in
a given configuration. This is a necessary step in designing an allocator. 

\begin{figure*}[ht]
    \centering
    \begin{subfigure}[t]{0.29\textwidth}
    \centering
    \includegraphics[scale=0.25]{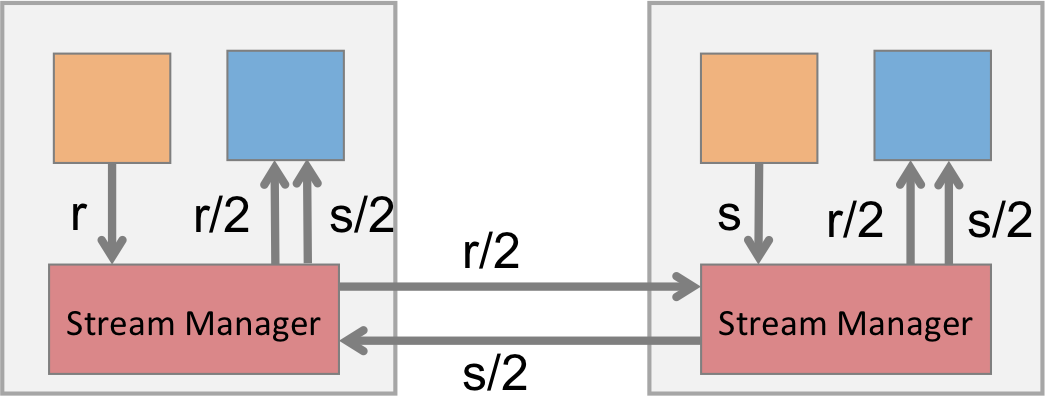}
    \caption{Modeling shuffle data rates}
    \label{fig:modeling-shuffle-data-rates}
    \end{subfigure}
    \begin{subfigure}[t]{0.29\textwidth}
    \centering
    \includegraphics[scale=0.30]{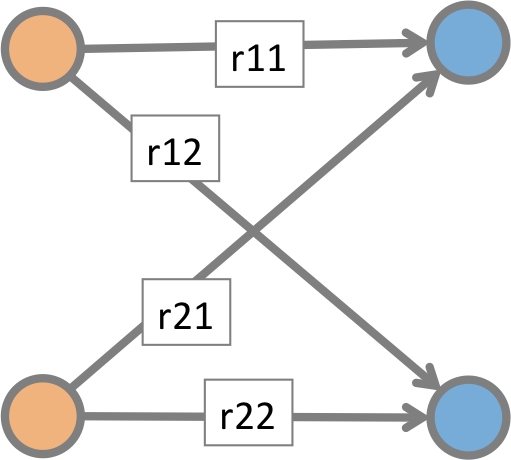}
    \caption{Modeling data grouping constraints}
    \label{fig:modeling-shuffle-constraints}
    \end{subfigure}
    \begin{subfigure}[t]{0.29\textwidth}
    \centering
    \includegraphics[scale=0.30]{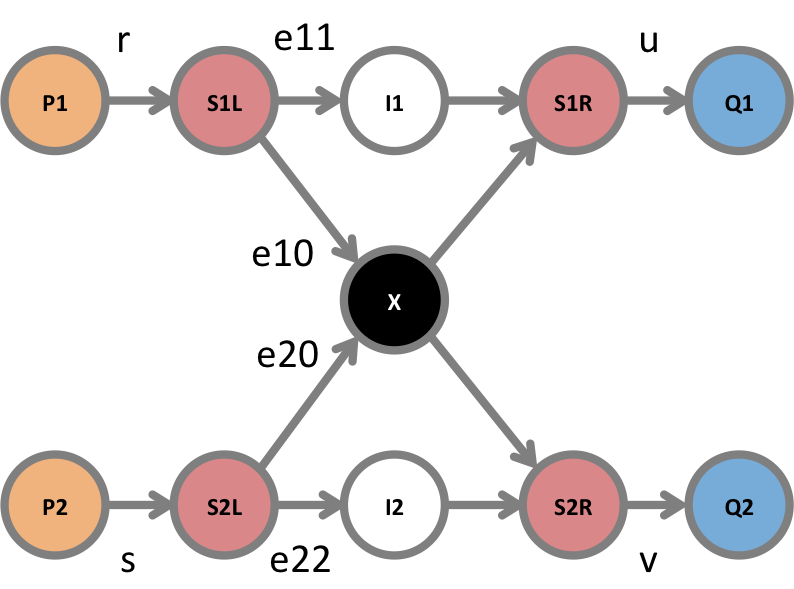}
    \caption{Modeling stream manager constraints}
    \label{fig:modeling-flow-constraints}
    \end{subfigure}
    \caption{Modeling data flow constraints to solve for processing rate of a configuration}
\end{figure*}

Our design of the data flow solver is based on the observation that a deployed
configuration can be seen as a physical {\em network} of DAG node instances
with resource constraints. Then, data flow can be analyzed by solving a network
flow problem. Node resource models and packing plan imply CPU and memory
constraints on node instances. Shuffling operators like fields-grouping or
shuffle-grouping and link capacities of container imply constraints on edges in
the network. An example of such modeling for WordCount is shown in fig.
\ref{fig:modeling-shuffle-data-rates}.

We model the network flow problem as a linear program with edge flow rates as
variables.  Node CPU models, container dimensions and link capacities are
expressed as linear constraints with a fairly standard formulation. Two sets of
constraints are unique to Heron configurations and critical for accurately
modeling the data flow: 1) data grouping constraints, and 2) stream manager
constraints. Both of those constraint sets encode the shuffling behavior of
edge operators connecting DAG nodes.

To understand the shuffling behavior and its translation into linear-program
constraints, consider the example of a WordCount configuration as shown in fig.
\ref{fig:modeling-shuffle-data-rates}. The DAG uses a fields-grouping shuffle
operator on the edge connecting the producer W to the consumer C.  This shuffle
operator is applied by the stream managers. Let us assume that a producer
instance outputs tuples at rate $r$ in steady state into the hosting stream
manager with uniform distribution over the key space.  Then, the stream manager
will split the data equally among the two consumers, passing $r/2$ to the local
consumer and remaining $r/2$ to the other container. The other stream manager
splits its producer's output rate $s$ similarly. Thus, in the resulting data
flow, while the configuration is processing a total of $r + s$, the stream
managers together are passing $1.5(r+s)$ in and out! The insight is that any
data tuple that crosses a container boundary passes through two stream
managers, thus costing twice the CPU resources for communication than a locally routed tuple. 
Accounting for intra- and inter-container data flow is the key to modeling 
communication cost and data flow correctly.

We encode data grouping behavior by constraining the edge set {$(u,v)$ for all
$v$} for any given node $u$. For example, in the network in
\ref{fig:modeling-shuffle-constraints}, we would write a constraint like
$r_{11} = r_{12}$. To encode stream manager constraints and correctly account
for CPU costs of intra- and inter-container tuples, we insert additional nodes
in the network by unfolding each stream manager into a left node that ingests
local tuples (e.g., S1L) and a right node (e.g., S1R) that emits local tuples.
We also add {\em internal routing nodes} (e.g., I1) for each container and a
single {\em switching node} X for the network, as shown in fig.
\ref{fig:modeling-flow-constraints}. Data routing locally from $SiL$ to $SiR$
passes through $Ii$. Data traversing container boundaries passes from $SiL$ to
$SjL$ ($i\neq j$) through X. This bifurcation of intra- and inter-container
data correctly accounts for the true CPU cost of communication.

Our linear-program based data flow solver acts as a complete performance
model for streaming workloads. It enables us to feed in an arbitrary
configuration and get an accurate prediction of tuple rate performance. In
doing so, it is able to precisely predict the flow rates within and across a
configuration using only DAG node models as application-specific input. As a
by-product, it also pin-points the rate-limiting parts of a configuration.
This gives the operator visibility into the specific configuration parameters
to tune in adjusting the target performance; it also gives the programmer
visibility into specific DAG nodes to optimize for most gain in performance.
For the case study of WordCount, the solver predicted 1050ktps for the true
value of 965ktps, amounting to less than 10\% of error. We evaluate the
prediction performance in more detail later.

\subsection{Model-based edge allocator}

We now present the design of an allocator that
can produce efficient configurations of a workload for any target data rate.  A
brute-force approach is to use a workload's performance model as a blackbox to
enumerate the performance of all possible configurations and then pick the best
one. However, as our previous discussion showed (sec. \ref{sec:overview}), this
approach is prohibitive
due to the sheer size of the configuration space.\footnote{Another equivalent
formulation is to cast the network flow problem as an integer linear program
(ILP) with binary assignment variables. With this formulation, even solving
WordCount over a few machines takes a few days.} Instead, we use the insights
gained from designing the data flow solver to design a closed-form algorithm
using DAG node models with linear time-complexity in the DAG size (nodes +
edges).

\begin{figure}[ht]
    \centering
    \includegraphics[scale=0.33]{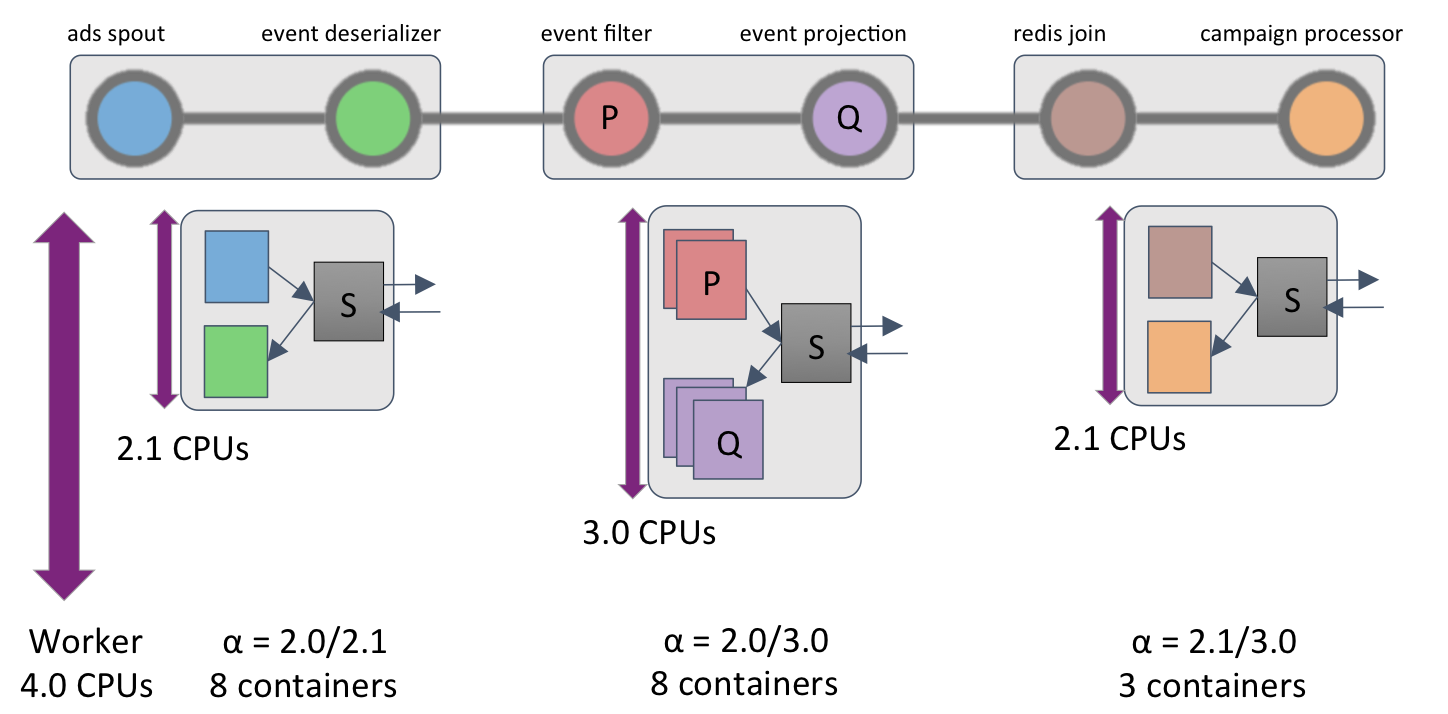}
    \caption{Provisioning using balanced containers}
    \label{fig:algorithm-illustration}
\end{figure}

Our allocation algorithm is based on the insight that a configuration is most
efficient for a certain data rate when all DAG nodes are rate matched, i.e.,
they are operating at their full capacity, {\em including the stream managers}.
Rate-matching DAG nodes to stream managers ensures that
computation and communication are also rate-matched.  Further, out of all
rate-matched configurations, the one with most data locality, i.e., least
cross-container data flow, will likely be the most efficient because it will
impose the least compute overhead of data communication.

We exploit the rate-matching and data-locality insights by allocating {\em
balanced edge containers} as basic units in the configuration. Consider the
example of the 6-node AdAnalytics DAG shown in fig.
\ref{fig:algorithm-illustration}. We first group nodes by edges alternate
edges.  Then, we compose a balanced container configuration for each edge, thus
co-locating communicating nodes for data locality. For an edge $P \rightarrow
Q$, a balanced container has $n_P$ and $n_Q$ instances of P and Q such that the
combined processing rates of those instances match up to each other and to the
processing rate of the stream manager, based on DAG node models. Further, in
this rate-matched configuration, we require that the stream manager is
operating at peak capacity by using a full CPU, so that the container is fully
utilized.

Rate-matching instances P and Q in a balanced container to stream manager S
poses a challenge: if P and Q are being allocated to process a rate R, then S
will be processing a rate more than R that depends on the final configuration.
The more the number of containers, the more S will be processing.  We solve
this cyclic dependency by assuming the worst-case for S.  We can show that as
containers increase in a configuration, S will need to pass a rate $4R$ in the
limit.

After composing balanced containers, the allocator configures the required
number of containers for each edge. Containers are replicated so that the
configuration can process the target rate at the input and the corresponding
rates (based on $\gamma$'s) on subsequent edges.  Additionally, the allocator
exposes a policy parameter for preferred container dimensions like
half-a-machine or third-of-a-machine. If the parameter is set, the allocator
scales down each balanced container before allocating copies of it. In the
AdAnalytics example, the balanced container for edge $P \rightarrow Q$ adds up
to a requirement of 3 CPUs at the rate-matching point. We emulate a scenario of
allocation half-sized containers that use 2 CPUs each. We scale each balanced
container with a factor $\alpha \leq 1$ to match this dimension. Finally, we
replicate $\alpha$-balanced-containers to the required count for target rate.
Optionally, container dimensions can also be searched over by passing in a set
of candidate container dimensions.

In the case of a general DAG with node degrees of more than 1, we generalize
edge allocation using a topological sorting of the DAG and successively
allocating critical paths based on compute cost. The time complexity of the
algorithm is $O(|V| + |E|)$ for nodes V and edges E in the DAG. We omit
the details for brevity.

%% file: implementation.tex
\section{Implementation}
\label{sec:implementation}

{\bf Prototype and execution speed:} We prototype \name as a stand-alone
auto-configuration and auto-scaling engine on top of Twitter Heron \cite{heron}.
Our implementation can use metrics from production workloads or 
test deployments. Training \name models does not require any code changes
to workloads or to the Heron runtime system. Once trained, models can predict
performance of a configuration in 10[msec] to 100[msec] time and the allocator
can produce a configuration for a target performance in 0.78[sec] on average. 
\\

\noindent {\bf Metrics:} The Heron runtime system provides a rich collection of
monitoring metrics for each node and stream manager in the topology. The main
load metrics we use are tuple rates on each edge in a deployed configuration.
The performance metrics per node instance we use are {\tt backpressure}, {\tt capacityutil}, {\tt
cputil}, {\tt memutil} and {\tt gctime}.  {\tt backpressure} is a measure of
time a node instance spends backlogged on data and slowing down upstream
components. {\tt capacityutil} is defined by the fraction of
time the node spent processing data as opposed to passively waiting for data. {\tt
cputil} and {\tt memutil} are traditional measures of CPU and memory
utilization. {\tt gctime} is a measure of time spent in Java
garbage collection.
\\

\noindent {\bf Modeling of CPU, I/O, memory, link, and node model drift detection:}
We have described \name node models in terms of CPU utilization, however,
nodes may not always be CPU-bound.  For example, Kafka data ingestion nodes in
our workloads make network calls to Kafka message brokers repeatedly, spending
significant time in network I/O. Modeling them purely through {\tt cputil} 
mispredicts their peak performance. 
We classify and train models for general nodes using the decision criteria in tab.
\ref{tab:all-metrics}. 

\newcommand{\tn}[1]{\textnormal{#1}}

\begin{table*}[t]
\centering
\begin{tabular}{|p{5.5cm}|p{4.0cm}|p{5.5cm}|}
\hline
\textbf{Metric combination} & \textbf{Implication} & \textbf{Treatment}\\
\hline
\small{$\tn{backpressure} > 0$} & \small{node saturated, miscalibrated} & \small{use caputil, cputil as saturating thresholds} \\
\hline
\small{$\tn{caputil} > 90\%$, $\tn{cputil} < 80\%$} & \small{node saturated, I/O-bound} & \small{normalize CPU model}\\
\hline
\small{$\tn{caputil} > 90\%$, $\tn{cputil} \geq 80\%$, gctime high} & \small{node saturated, memory-bound} & \small{allocate more memory and retrain}\\
\hline
\small{$\tn{caputil} > 90\%$, $\tn{cputil} \geq 80\%$, gctime low} & \small{node saturated, CPU-bound} & \small{use default CPU model}\\
\hline
\end{tabular}
\caption{Decision criteria for training CPU-bound and I/O bound node models}
\label{tab:all-metrics}
\end{table*}

If we detect backpressure in an instance at runtime, we treat it as a sign
of performance saturation.  We learn the saturation points for different
resources for that node and include that as part of the node model. 
If backpressure is caused in a \name allocation, it signals node model error 
either due to noise or model drift. 
In either case, we trigger re-training of that node's model.

If we classify a node to be I/O-bound, we normalize its CPU model such that
{\tt cputil} saturates with {\tt capacitytil}. This hides some inefficiency in
terms of CPU allocation to the node but ensures that the resulting
configuration is feasible.

\begin{figure}[!h]
    \centering
    \includegraphics[scale=0.28]{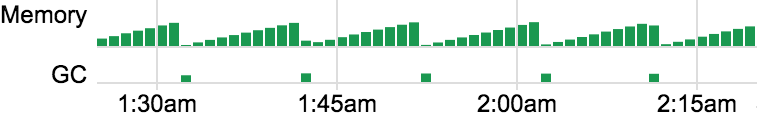}
    \caption{A typical memory sawtooth with garbage collection spikes}
    \label{fig:memory-sawtooth}
\end{figure}

We learn the memory requirement of a node by combining {\tt memutil} and
{\tt gctime} metrics. We notice a typical sawtooth memory utilization pattern
as shown in fig. \ref{fig:memory-sawtooth}. A node instance continues to draw
memory until the JVM environment has used up all available memory. At that
point, the garbage collector (GC) is triggered, immediately bringing down
memory usage substantially. This usage pattern results in {\tt memutil} being
an oscillating timeseries that can lead to over-estimation of
average memory requirement. We filter down the {\tt memutil} timeseries to
samples right after GC trigger points to glean the true memory requirement of a
node. Further, with this filtered data, we model memory as a function of
tuple rate, much like CPU. The intuition is that many streaming
operations maintain key-value windowed data-structures like hashmaps that grow
in proportion to the key-space mapped to a node instance. As a node is
parallelized more, it receives a smaller subspace of keys on any instance and
therefore consumes less memory per instance. The memory-scaling model is
incorporated into the allocator as a resource constraint when composing a
balanced container. This modeling allows us to find configurations that can
work with tight memory bounds.

Network utilization is modeled by simply using bytes accounting metrics and
using them as scaling factors on tuple rates. Link rates are resource constraints on containers similar to memory. 
\\

\noindent {\bf End-to-end model calibration, noise margin and drift detection:}
Some noise is inherent in the models due to variations in performance caused by
system effects like kernel scheduling, caching and I/O. In addition, there is a
systematic sampling bias effect that can cause errors in the models.
Any node is sampled for CPU $\sim$ tuple-rate relation in a limited dynamic
range of CPU utilization depending on the configuration and the load range. For
example, if a configuration over-provisions a node, each instance might only
operate up to a peak {\tt cputil} of 20\%. Using a model from this data in
predicting the tuple-rate at {\tt cputil} of 80\% succumbs to errors. 

We handle the sampling bias problem with two safeguards. First, we use the
trained model to predict back the performance of sampled configurations to
compute the training error.  We use this feedback loop to tune an internal
over-provisioning factor to eliminate training error. For example, if we
predict 1050ktps for a configuration measuring to 965ktps, we know we are
over-predicting performance by 9\%. Then, we set the over-provisioning factor
to 1.09, so that the allocator produces configurations for 1.09 * target-rate.
Second, as we deploy \name-generated configurations, we keep pooling metrics
and improve model performance. The reason is that \name configurations deploy
all nodes with rate-matching, so they operate close to their capacity. This
pushes node instances into higher and more important ranges of resource
utilization, providing us better sampling.

Finally, when the training error becomes bigger than a certain threshold, we
declare model drift and trigger retraining.

%% file: eval.tex
\squeezeup
\begin{figure}[ht]
    \centering
    \includegraphics[scale=0.25]{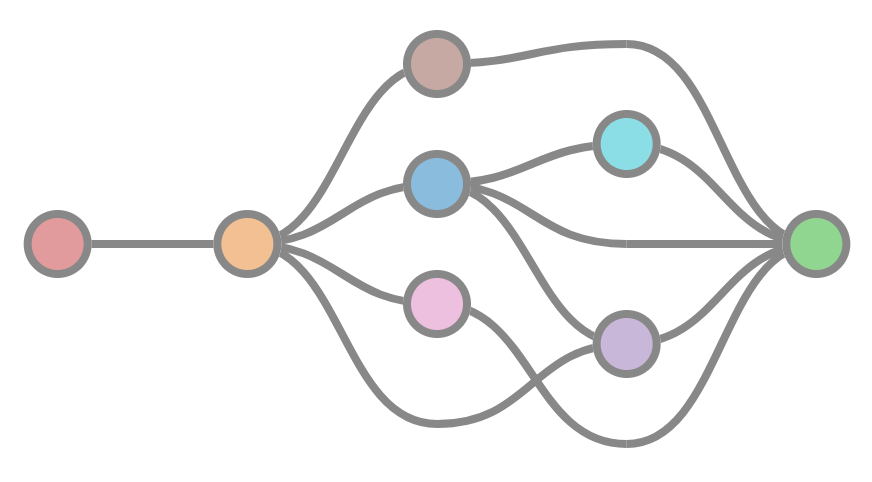}
    \caption{A mobile network user-analytics logical DAG}
    \label{fig:uhanaapp}
\end{figure}
\squeezeup

\begin{figure*}[ht]
    \centering
    \begin{subfigure}[t]{0.32\textwidth}
    \centering
    \includegraphics[scale=0.2]{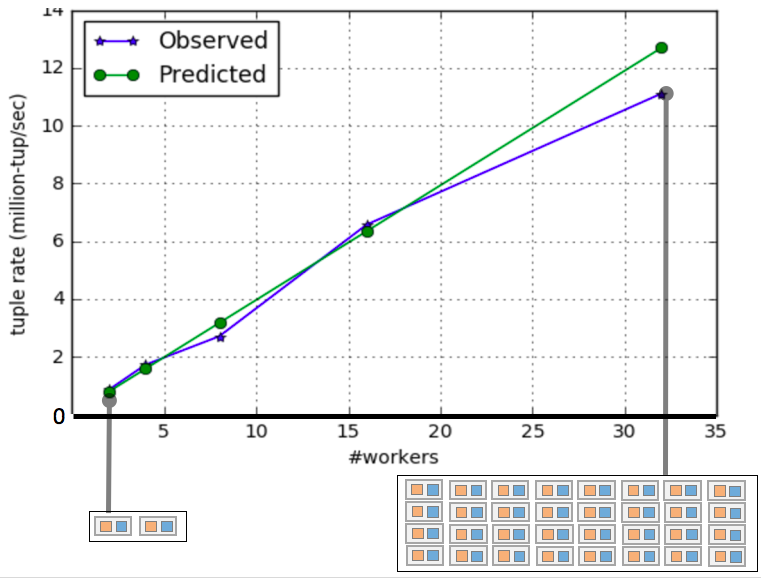}
    \caption{WordCount scaling performance prediction}
    \label{fig:scale}
    \end{subfigure}
    \begin{subfigure}[t]{0.32\textwidth}
    \centering
    \includegraphics[scale=0.2]{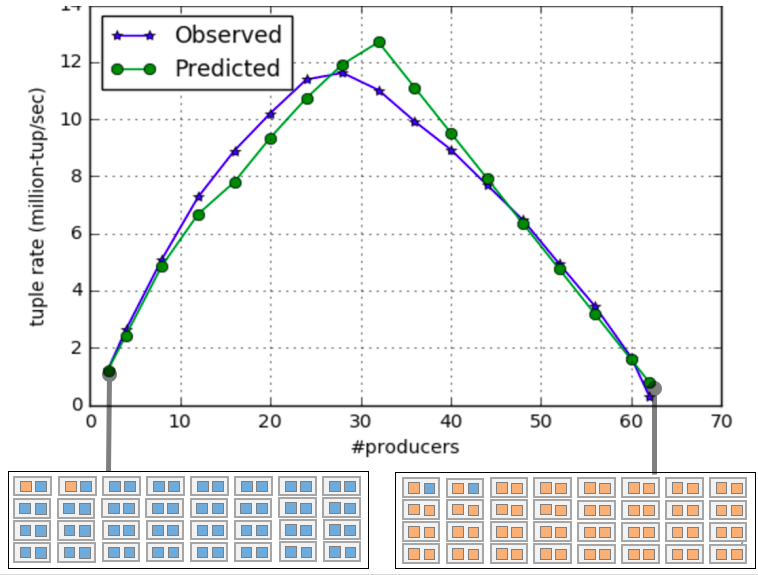}
    \caption{WordCount parallelism variance performance prediction}
    \label{fig:alternate}
    \end{subfigure}
    \begin{subfigure}[t]{0.32\textwidth}
    \centering
    \includegraphics[scale=0.3]{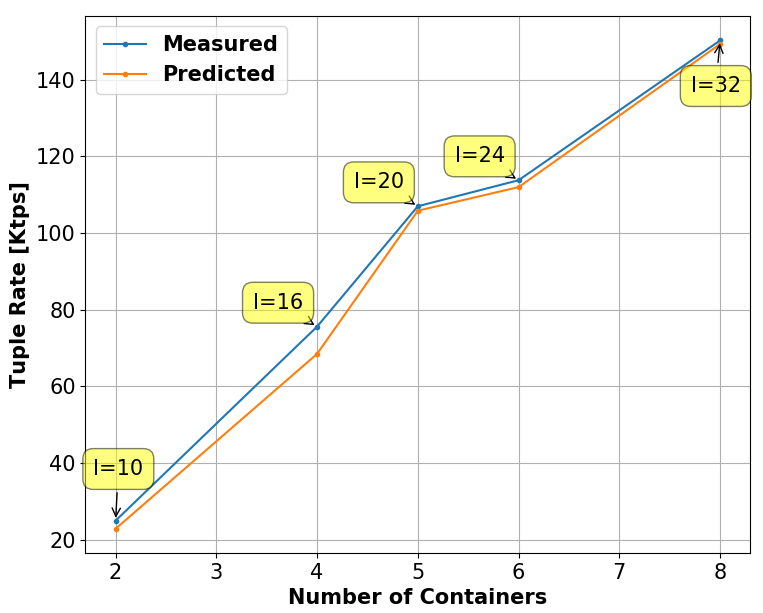}
    \caption{Performance prediction for the mobile network user-analytics workload}
    \label{fig:uhannaapp-pred}
    \end{subfigure}
    \caption{\name prediction performance for AdAnalytics and WordCount}
\end{figure*}

\begin{figure*}[!h]
    \centering
    \begin{subfigure}[t]{0.32\textwidth}
    \centering
    \includegraphics[scale=0.28]{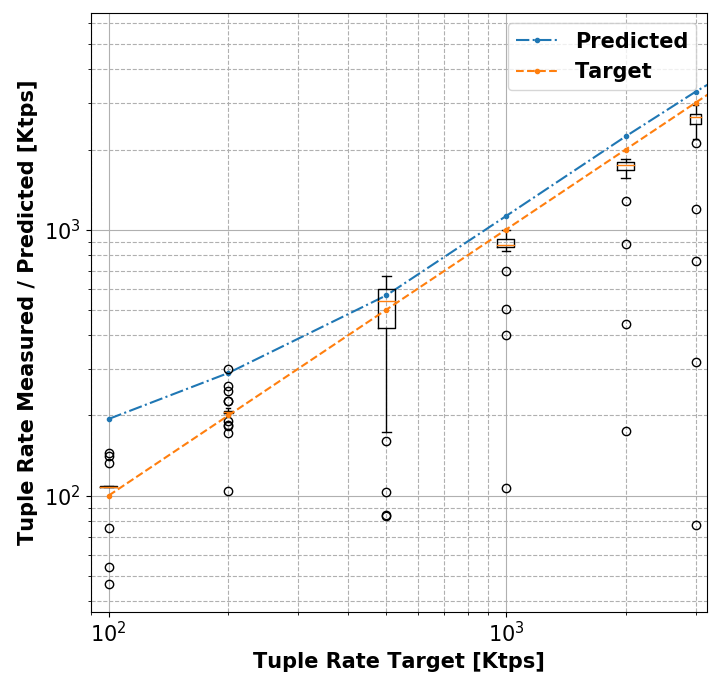}
    \caption{Ad analytics performance using \name allocator}
    \label{fig:yahoo-res}
    \end{subfigure}
    \begin{subfigure}[t]{0.32\textwidth}
    \centering
    \includegraphics[scale=0.28]{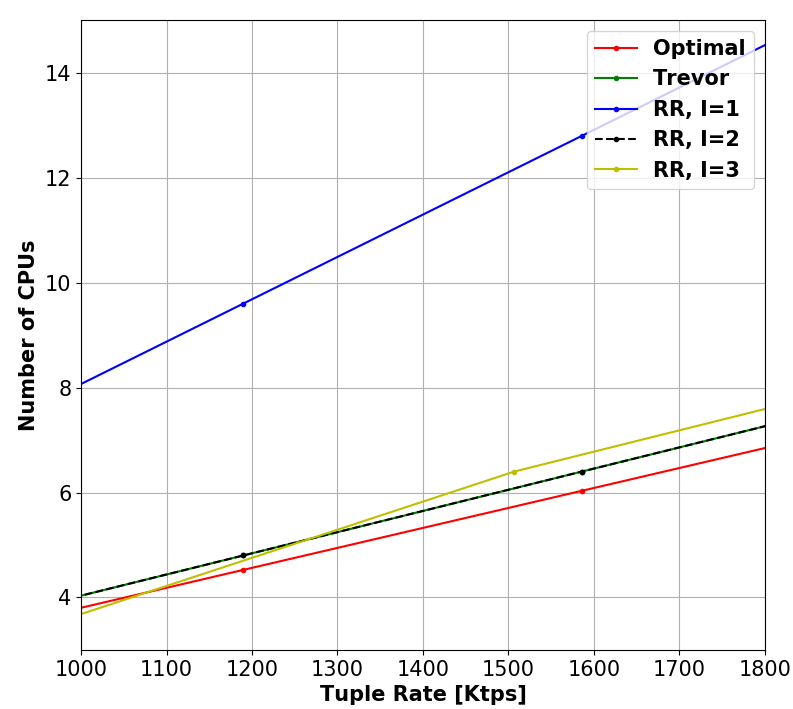}
    \caption{WordCount pipeline CPU usage for different configurations}
    \label{fig:cpu_util_word}
    \end{subfigure}
    \begin{subfigure}[t]{0.32\textwidth}
    \centering
    \includegraphics[scale=0.28]{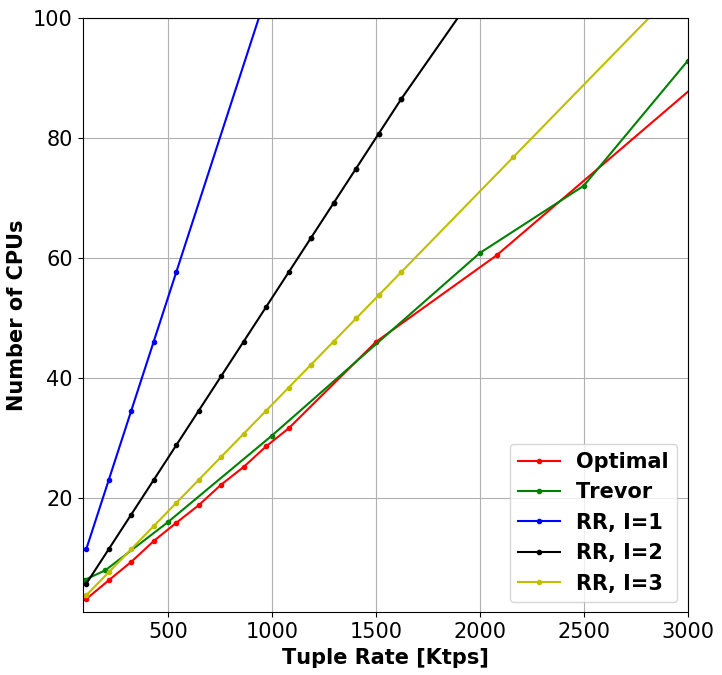}
    \caption{Ad analytics pipeline CPU usage for different configurations}
    \label{fig:cpu_util_yahoo}
    \end{subfigure}
    \caption{\name evaluation results}
\end{figure*}

\section{Evaluation}

In this section, we evaluate two aspects of \name's performance: (1) its predictive models for
components in a DAG, and (2) its allocator algorithm that returns an optimal allocation for a
target input rate.

\subsection{Workloads and test setup}
\label{sec:workloads}

We evaluate \name on three workloads: the WordCount workload in 
fig.~\ref{fig:wordcount}, the AdAnalytics workload~\cite{yahoo-streaming-benchmark} in fig.~\ref{fig:yahooapp},
and a mobile network user-analytics workload in fig.~\ref{fig:uhanaapp}. 
All three workloads represent real world topologies with increasing degrees of complexity.

For our experiments, we use clusters with 4-CPU VMs, and deploy them on Heron 0.14.3.
At the source of the DAG we deploy a cluster running Kafka~\cite{kafka} as input. 
To sweep over a range of input rates, we run a ConsoleProudcer throttled to the specified rates,
with appropriate hold times for each rate. We set our sampling interval to 10 seconds, swept
over a wide range of inputs so the models generalize, and profiled the system over a span of 
20 minutes. In a production setting, a profiling over the course of day should be sufficient to
capture natural variation in load patterns; the models can then be refined in subsequent days,
with no change to the deployment setup.

\subsection{Model Accuracy}
\label{sec:modeling-accuracy}

\begin{table}[ht]
\centering
\footnotesize
\begin{tabular}{|p{2.3cm}|R{0.65cm}|p{1cm}|R{0.65cm}|p{1cm}|}
\hline
\textbf{Instance} & \textbf{CPU $R^2$} & \textbf{CPU Range} & \textbf{Cap $R^2$} & \textbf{Cap Range} \\
\hline
\footnotesize{ads} & \footnotesize{0.597} & \footnotesize{0.10-0.32} & \footnotesize{0.741} & \footnotesize{0.90-1.00} \\
\hline
\footnotesize{event deserializer} & \footnotesize{0.933} & \footnotesize{0.10-1.10} & \footnotesize{0.956} & \footnotesize{0.10-0.90}\\
\hline
\footnotesize{campaign processor} & \footnotesize{0.763} & \footnotesize{0.01-0.05} & \footnotesize{0.500} & \footnotesize{0.01-0.25}\\
\hline
\footnotesize{event projection} & \footnotesize{0.715} & \footnotesize{0.03-0.13} & \footnotesize{0.998} & \footnotesize{0.01-0.15}\\
\hline
\footnotesize{event filter} & \footnotesize{0.892} & \footnotesize{0.10-0.40} & \footnotesize{0.990} & \footnotesize{0.01-0.17}\\
\hline
\footnotesize{redis join} & \footnotesize{0.758} & \footnotesize{0.01-0.16} & \footnotesize{NA} & \footnotesize{NA}\\
\hline
\footnotesize{stream manager} & \footnotesize{0.797} & \footnotesize{0.01-0.26} & \footnotesize{NA} & \small{NA}\\
\hline
\end{tabular}
\caption{$R^2$ for Ad Analytics model fit}
\label{table:modeling-mse}
\end{table}

We first demonstrate the accuracy of our instance modeling techniques on the AdAnalytics DAG
in fig.~\ref{fig:models_yahoo}, a more complex DAG than WordCount. The $R^2$ numbers
for both the CPU and capacity linear models, shown in tab.~\ref{table:modeling-mse}, indicate
that CPU utilization and DAG node capacity have a strong linear relationship with input rate.

We demonstrate \name's predictive ability under different scaling behaviors with the WordCount workload. 
In the first test, we begin with a container of one producer and one consumer, and incrementally
scale the number of containers and machines. We compare the measured performance with our predicted
performance, as shown in fig.~\ref{fig:scale}. \name accurately predicts the
performance with an error of at most 10\% of the measured result. 

In the second test we evaluate an allocation in which we have 64 instances of either the producer or the consumer.
We start with 2 producer and 62 consumers and incrementally shift producers to consumers until we have 62 producers and 2 consumer.
The results are presented in fig.~\ref{fig:alternate}. \name correctly predicts the performance as we change the ratio of parallelism between producers and consumers. We see that \name predicts the optimal configuration to be very close to the true optimal configuration.

In fig.~\ref{fig:uhannaapp-pred}, we show that our predictions remain accurate on complex
nonlinear DAGs, such as the mobile network user-analytics DAG. 
\name again predicts with a 10\% error from the measured result.

The runtime of \name's prediction system varies based on the size of the DAG. In our example
topologies, predictions ran between 10 and 100 millseconds.

\subsection{Allocation Efficiency}
\label{sec:allocation-efficiency}

We evaluate \name's allocator algorithm by showing we can configure the AdAnalytics workload to hit 
an increasing range of tuple rates.
As we saw in the previous section, we can correctly predict the performance of an allocation with an error rate of 10\%. To account for the 10\% error rate we overprovision our workload by 10\%-20\%. For instance, if we want to hit a rate of 1Mtps we run our allocator for 1.15Mtps.

The performance of our allocator's configuration algorithm at different input rates is shown in
fig.~\ref{fig:yahoo-res}. The predicted line shows the rate \name predicted for the optimal
allocation generated, and the target line shows the target rate we configured for.
We can see that we can generally correctly configure using the allocator by maintaining the 10\% over prediction rate.

To measure \name's resource utilization, we compare its allocation with optimal resource 
utilizations
and different paths of round robin allocation scaling in fig.~\ref{fig:cpu_util_word} for WordCount and 
fig.~\ref{fig:cpu_util_yahoo} for AdAnalytics.

We calculate the optimal line by placing all instances of the DAG in a single container with unlimited 
resources, and we remove the resource limitation on the stream manager. We increase parallelism and see how many 
CPUs where required to hit a given tuple rate. For round robin allocation, we decide how many instances are placed in each
container, and we scale by increasing the number of containers and instances while maintaining this ratio 
(donated by I in the graph).

In Figure-\ref{fig:cpu_util_word} we can see that round robin performs quite well for the simple WordCount DAG. But once we move to a more complex DAG such as AdAnalytics in Figure~\ref{fig:cpu_util_yahoo}, \name
beats all round robin allocations in CPU and tracks the optimal CPU usage quite well, differing by 10\% at most from the optimal.

\name's allocator also perform well in terms of the run time required to calculate an allocation. The balanced container's
algorithm is a closed form algorithm dependent only on the graph complexity and not the target input rate. For AdAnalytics
calculating an allocation took 0.7811 seconds on average across all rates.

%% file: related.tex
\section{Related work}
\label{sec:related}

The work closest to ours that addresses auto-tuning and auto-scaling
in stream processing systems is Dhalion\cite{dhalion}. Dhalion is a 
self-regulating system built on the Heron\cite{heron} platform, which 
enables auto-scaling and auto-tuning through a stepwise iteration over
allocations during runtime, incrementally reacting to bottlenecks by adjusting parallelism until
the topology is stable.

\name on the other hand requires no runtime iteration, and can reach a given
target performance with a closed form algorithm, allowing for faster 
tuning and scaling than Dhalion. Additionally, \name achieves a CPU efficient
allocation while Dhalion will maintain inefficiencies that are present in the 
initial allocation it is regulating.

Other works for auto-scaling for stream processing have previously been proposed, such as
Elastic Auto-Parallelization\cite{gedik2014elastic} and DRS\cite{fu2015drs}. These methods, however,
are not designed to scale and tune the topology based on the input rates, as with \name.

There has been a lot of prior work on auto-tuning in the batch processing world, such as Hemingway\cite{hemingway} and ROPE\cite{rope}. These works are based on multiple techniques including modeling compute and communication in relation to tuning parameters and identifying bottlenecks 
Many of these works 
utilize learning techniques from past data \cite{mr-workload-suites,cross-industry-mr-workloads}. 

A few works such as Quasar\cite{quasar}, CherryPick\cite{cherrypick}, Throughput-Scheduler\cite{throughput-scheduler}, Proactive ReOptimization\cite{proactive-reoptimization} and Starfish\cite{starfish}, 
like \name, try to optimize for a goal, based on performance.
Yet both papers are based on a black box model of the entire application, which requires
a search involving many iterations over a large search space. As we have shown with \name, this
might result in inefficient stream processing applications.

The techniques presented in the batch processing papers target a different set of applications 
than \name does, and are not directly applicable to stream processing. More over batch 
processing systems are rescheduled periodically, and can be modified between 
each run. Stream processing systems are scheduled once and then regulated.

%% file: conclusion.tex
\section{Conclusion}
\label{sec:conclusion}
In this paper, we introduced \name, an auto-tuning 
system for stream processing pipelines. We demonstrated the difficulty
of tuning streaming pipelines to a given performance, and showed
that this can result in both inefficiencies in resource utilization and over-provisioning. We demonstrated how \name uses effective node modeling
to learn and predict the pipeline performance with a 10\% error rate. We showed
how \name's predictive models can be used to design the \name allocator, which 
produces an optimal allocation for the topology, given a target rate.
Additionally, we have shown that the \name allocator can---in less than 1 second---produce an allocation which 
achieves the desired target rate.
